\documentclass[12pt,preprint]{emulateapj}
\usepackage{psfig}

\def\H0{{\rm ~km~s^{-1}~Mpc^{-1}}}

\shorttitle{GRB 030115}
\shortauthors{A.J. Levan et al.}

\begin{document}



\newcommand\hyperz{{\sc Hyperz}}
\newcommand\BPZ{{\sc BPZ}}
\newcommand\IRAF{{\sc IRAF}}
\newcommand\ORAC{{\sc ORAC-DR}}
\newcommand\gcn{GCN circ.}
\newcommand\etal{{\it et al.}}
\newcommand\delt{{$\Delta t$}}

\title{Infrared and Optical Observations of GRB~030115 and its Extremely Red Host
Galaxy: Implications for Dark Bursts$^1$}

\author{
Andrew~Levan\altaffilmark{2,3},
Andrew~Fruchter\altaffilmark{3},
James~Rhoads\altaffilmark{3},
Bahram Mobasher\altaffilmark{3},
Nial Tanvir\altaffilmark{4},
Javier Gorosabel\altaffilmark{3,5},
Evert~Rol\altaffilmark{2,6},
Chryssa Kouveliotou\altaffilmark{7},
Ian~Dell'Antonio\altaffilmark{8,9},
Michael~Merrill\altaffilmark{10},
Eddie Bergeron\altaffilmark{3},
Jos\'e Mar\'{\i}a Castro Cer\'on\altaffilmark{3},
Nicola Masetti\altaffilmark{10},
Paul Vreeswijk\altaffilmark{11},
Angelo Antonelli\altaffilmark{12},
David Bersier\altaffilmark{3},
Alberto Castro-Tirado\altaffilmark{5},
Johan Fynbo\altaffilmark{13},
Peter Garnavich\altaffilmark{14},
Stephen Holland\altaffilmark{15}
Jens Hjorth\altaffilmark{16},
Peter Nugent\altaffilmark{17},
Elena Pian\altaffilmark{18},
Alain Smette\altaffilmark{19},
Bjarne Thomsen\altaffilmark{11},
Stephen E. Thorsett\altaffilmark{20},
Ralph Wijers\altaffilmark{6} \\}

\altaffiltext{1}{Based in part on observations made with the NASA/ESA {\it Hubble
Space Telescope,} obtained at the Space Telescope Science Institute,
operated by the Association of Universities for Research in
Astronomy, Inc., under NASA contract NAS 5-26555. These observations
are associated with programs 9074 and 9405.}

\altaffiltext{2}{Department of Physics and Astronomy, University of 
Leicester, University Road, Leicester, LE1 7RH, UK} 
\altaffiltext{3}{Space Telescope Science Institute, 3700 San Martin 7 
Drive, Baltimore, MD21218, USA} 
\altaffiltext{4}{Centre for Astrophysics Research, University of Hertfordshire, College  
Lane, Hatfield, Hertfordshire, AL10 9AB, UK} 
\altaffiltext{5}{Instituto de Astrof\'{\i}sica de Andaluc\'{\i}a (IAA-CSIC), 
    P.O. Box 03004, E-18080 Granada, Spain.} 
\altaffiltext{6}{Astronomical Institute, University of Amsterdam, 
Kruislann 403, 1098 SJ Amsterdam, The Netherlands} 
\altaffiltext{7}{NASA/MSFC, NSSTC, XD-12, 320 Sparkman Drive, Huntsville, AL 35805, USA} 
\altaffiltext{8}{Physics Department, Brown University, Providence, RI 
02912, USA} 
\altaffiltext{9}{National Optical Astronomy Observatory, P.O. Box 
26732, Tucson, AZ 85726-6732, USA} 
\altaffiltext{10}{Istituto di Astrofisica Spaziale e Fisica Cosmica - 
Sezione di Bologna, CNR, via Gobetti 101, 40129, Bologna, Italy} 
\altaffiltext{11}{European Southern Observatory, Casilla 19001, Santiago 
19, Chile}
\altaffiltext{12}{Osservatorio Astronomico di Roma, Via Frascati 33, Monteporzio I-00040, Italy.} 
\altaffiltext{13}{Department of Physics and Astronomy, Aarhus University, Ny Munkegade, 
DK-8000, Aarhus, Denmark} 
\altaffiltext{14}{Department of Physics, University of Notre Dame, 
Notre Dame, IN 46556-5670, USA} 
\altaffiltext{15}{US Naval Observatory, Flagstaff Station, P.O. Box 1149, Flagstaff, AZ 
86002, USA} 
\altaffiltext{16}{Astronomical Observatory, University of Copenhagen, Juliane Maries Vej 
30, DK-2100, Copenhagen, Denmark} 
\altaffiltext{17}{Lawrence Berkeley National Laboratory, 1 Cyclotron Road, 
Berkeley, CA 94720, USA} 
\altaffiltext{18}{Osservatorio Astronomico di Trieste, Via G.B. Tiepolo 11, 34131 Trieste, Italy} 
\altaffiltext{19}{Institut d'Astrophysique et de G\'eophysique, 
Universit\'e de Li\`ege, Avenue de Cointe, 5, B-4000 Li\`ege, Belgium} 
\altaffiltext{20}{Department of Astronomy and Astrophysics,  
University of California, 1156 High Street, Santa Cruz, CA 95064, USA}

\begin{abstract} 
We present near-infrared (nIR) and optical observations of the afterglow of 
GRB~030115.  Discovered in an infrared search at Kitt Peak 5 hours after the burst trigger, 
this afterglow is amongst the faintest observed in the R-band at an early 
epoch, and exhibits very red colors, with $R-K\approx 6$.  The magnitude of  
the optical afterglow of GRB~030115 is fainter than many upper limits for other bursts, 
suggesting that without early nIR  
observations it would 
have been classified as a ``dark'' burst. Both the color and optical magnitude 
of the afterglow are likely due to dust extinction and indicate that at least 
some optical afterglows are very faint due to dust along the line of sight.  
Multicolor {\it Hubble Space Telescope} observations were also taken of  
the host galaxy and the surrounding 
field. Photometric redshifts imply that the host, and a substantial number of 
faint galaxies in the field are at $z \sim 2.5$.  The overdensity of galaxies 
is sufficiently great that GRB~030115 
may have occurred in a rich high-redshift cluster. The host galaxy shows 
extremely red colors (R-K=5) and is the first GRB host to be classified as  
an Extreme Red Object (ERO). Some of the galaxies surrounding the host also show 
very red colors, while the majority of the cluster are much bluer, indicating ongoing 
unobscured star formation. As it is thought that much of high redshift star formation 
occurs in highly obscured environments it may be that GRB 030115 represents 
a transition object, between the relatively unobscured afterglows seen to date and 
a population which are very heavily extinguished, even in the nIR. 
 
\end{abstract}

\section{Introduction}  
Most ground-based GRB afterglow searches are being pursued first in optical wavelengths.  
These have relied on either comparing 
a newly acquired image to an archival image (normally the Digitized 
Sky Survey) and locating any new sources, or on performing multi-epoch 
observations and identifying any variable sources within the GRB error 
box. In many cases both approaches have been used. Such searches have 
resulted in the discovery of approximately 60 GRB afterglows with 
about 50\% and 35\% being discovered, respectively, by each of the 
methods, and $\sim15\%$ being discovered by a combination thereof. One 
GRB afterglow was found by a color-color technique, (Rhoads 2001; 
Gorosabel \etal\ 2002), exploiting the ability of multi-band 
observations to distinguish blackbody stellar spectra from the 
power-law spectra displayed by GRB afterglows.  
 
The advent of small 
robotic telescopes that can slew automatically in response to GRB alert 
notices (e.g., ROTSE) and larger telescopes that can easily be 
remotely operated (e.g., the Palomar 48~inch) has increased the 
discovery rate of young and bright afterglows. 
For example, the afterglows of GRB~021004 and GRB~021211 were 
discovered only minutes after the GRB trigger (Fox 2002; Fox \etal\ 
2003a,b). These rapid-response campaigns have also substantially 
decreased the fraction of bursts that have no identified optical 
counterparts: the so called ``dark bursts.'' Of the GRBs well localised so 
far by the Wide X-ray Monitor (WXM) on {\it HETE-2,} the optical 
counterpart discovery percentage is $\sim 60\%$, while for bursts 
localized by the Soft X-Ray Camera (SXC) the discovery rate is $\sim 
90\%$, (largely due to the smaller error boxes associated with
SXC positions - see for example, Lamb \etal\ 2003). There are evidently 
possible selection effects: relatively bright and soft bursts are more 
easily localised by the SXC, and these smaller error boxes are easier to 
search to deep limits. Indeed, although {\it Swift} now allows 
routine $\sim 3$ arcminute error boxes to be routinely 
circulated its increased sensitivity means that it locates fainter
bursts, and the recovery rate of optical afterglows is lower than for the SXC.
Nevertheless the dark burst fraction does appear to have decreased
significantly from the $\geq 70\%$ that was typical during the
first few years of GRB observations, when slow and less
accurate positions made it more difficult to locate afterglows,
even for relatively bright bursts (see e.g. Fynbo \etal\ 2001; Lazzati \etal 2002).

The spectral energy distribution (SED) of a GRB afterglow is best 
described by a set of connected power-laws with breaks due to 
synchrotron self-absorption, the peak energy of the electrons, and the 
time-dependent cooling of the fireball (Sari, Piran \& Narayan 
1998). In theory, knowledge of these break frequencies allows us to  
reconstruct details of the afterglow (as has been done for 
a number of well sampled GRB afterglows to-date, e.g., by Panaitescu 
\& Kumar 2001). Theoretical models are also useful for studying GRBs 
with no optical afterglows, where observations in some bands (such as 
X-rays and radio) can be used to place limits on the expected optical 
flux.  In a prototypical dark burst, such as GRB~970828 (Groot \etal\ 
1998; Djorgovski \etal\ 2001), extrapolation of the X-ray flux, 
assuming a synchrotron spectrum, led to the conclusion that the burst 
should have been seen in the optical.  Conversely, Fynbo \etal\ (2001) 
have shown that the classification of many bursts as dark may simply 
represent the failure to search either early or deep enough. This 
interpretation is supported by the detection of the faint afterglow of GRB~020124 
(Berger \etal\ 2003) and the fast fading afterglow of GRB~021211 
(Fox \etal\ 2003a; Crew \etal\ 2003), which would not have been detected but for early, 
deep observations. Although it is also true that the refinement of afterglow
locations via X-ray observations does allow for very deep pointed observations, 
and thus enables faint afterglows to be located. This is now commonly 
the case for {\it Swift} bursts. 
 
For the relatively small number of bursts that are still apparently 
``dark'' despite deep and early optical limits, a number of plausible 
explanations have been widely discussed including
i) bursts which occur at high$-z$ where the Lyman break has moved
through the optical bands ii) GRBs which originate from behind significant 
obscuring columns, significantly attenuating the optical light, iii) bursts
which are intrinsically faint (i.e. are spectrally similar to bright) afterglows but 
are a factor 10-100 fainter and iv) fade very rapidly so that, at the time
of the first optical observations they have faded below the detection limit. 
Observations of afterglows to date, including recent observations of 
{\it Swift} bursts indicate that all of the above explanations may apply
with some bursts originating from beyond z=5 (e.g. Jakobsson et al. 2005; 
Haislip et al. 2005; Kawai et al. 2005), a fraction being apparently
obscured (e.g. De Pasquale et al. 2005; Watson et al. 2005).
While the first two scenarios above 
have been used to motivate afterglow searches in the near-infrared, 
locating afterglows from bursts akin to the third and fourth options 
can be equally well done by early, deep optical observations.
 
As the connection between long-duration 
GRBs and supernovae is now secure 
(Hjorth \etal\ 2003; Stanek \etal\ 2003; Zeh, Klose \& Hartmann 2004), we expect GRBs to point 
to locations of star formation throughout the Universe. The determination of which of the  
explanations above is the dominant origin of dark GRBs may thus  
shed light into the processes of star formation and its evolution.  
For example were many dark GRBs genuinely at high redshifts, they would point to strong star formation even at very early ages in the Universe, while if most are actually in obscured  
environments at lower redshifts they would support models 
whereby much of the star formation 
in the Universe is obscured. Moreover, the type of their host galaxies,  
their environments and relative frequencies may also indicate where much 
of the star formation occurs.   
 
Here we present the nIR discovery of 
the afterglow of GRB~030115. This burst was heavily reddened with 
respect to a model power law SED and, in the optical, was the faintest 
GRB afterglow ever observed at similar epochs post-burst.  We show 
that its host galaxy lies at moderate redshift and has very red colors 
(R$-$K $\sim 5$). It is therefore likely that GRB~030115 was reddened 
due to high local extinction. Furthermore there is a significant overdensity 
of galaxies at the redshift of the GRB host possibly associated with the latter; this putative galaxy cluster is a mix of young, blue, star forming systems
and red systems which are either dusty starburst systems or contain an older
stellar population. 
This cluster, if confirmed, is the first associated with a GRB, and one  
of very few known at $z>2$. 
 
Throughout this paper we assume a $\Lambda CDM$ cosmology with 
$\Omega_{\Lambda} = 0.73$, $\Omega_M = 0.27$  
and $H_0 = 72$~km~s$^{-1}$~Mpc$^{-1}$. 

\section{Observations}  
 
GRB~030115 was detected by the {\it HETE-2} satellite on 
2003 January 15 at 02:44 UT, with a duration of $t_{90} = 17.8$ seconds (Kawai  
\etal\ 2003). The burst, classified as an X-ray Rich GRB\footnote{see 
http://space.mit.edu/HETE/Bursts/Data/}, was localised by both the  
WXM and SXC 
instruments with error circles of 5 and 2 arcmin radius, respectively. 
Early optical observations were acquired by several groups using 
1-2~m class telescopes. Comparing these images with the Digitized Sky Survey 
revealed no new objects, placing an early (\delt$=t_{\rm obs}-t_0 \approx2$ hour)  
limit on the 
magnitude of the afterglow at $R=21$ (Castro Tirado \etal\ 2003; Blake \etal\ 
2003). We obtained data from the Kitt Peak National Observatory (KPNO)  
5 and 9 hours after the GRB in the infrared 
(J,H,K) bands, comparison of which revealed a fading IR source in all filters 
(see \S\ref{sec:ir}). Following our announcement of the IR counterpart, a re-inspection 
of early optical images obtained at \delt =2 hours revealed a marginal optical 
detection (Masetti {\etal\  2003a). Further IR observations were obtained by Vrba 
\etal\ (2003),  Kato \etal\ (2003) and Dullighan et al. (2004) and
confirmed the fading nature of the object, 
thus securing its association with GRB~030115. Early radio observations at 8.46 
GHz (Berger \& Frail, 2003) failed to detect the afterglow candidate at \delt =5  
hours, specifically recording a flux density  
of $84 \pm 62 \mu$Jy. However, further observations at \delt =55 hours 
detected a weak source coincident with that of the afterglow, with 
a flux level of $94 \pm 22 \mu$Jy (Frail \& Berger 2003). Radio observations 
were also undertaken at the Westerbork Synthesis Radio Telescope (WSRT) at 3 
epochs between 1.5 and 12 days post-burst. Combining these data provided a $\sim 
4 \sigma$ detection of the radio transient at 4.9~GHz (Rol \& Wijers 2003). 
Sub-millimeter observations were pursued with the SCUBA array on the James 
Clark Maxwell Telescope (JCMT) at 850~$\mu$m (Hoge \etal\ 2003) and the 
Max-Planck Millimeter Bolometer (MAMBO) and IRAM 30-m telescope at 1.2 mm 
(Bertoldi \etal\ 2003). These placed 3-$\sigma$ limits at 850~$\mu$m and 
1.2~mm of 6~mJy and 3~mJy, respectively. While these observations are relatively 
deep (for afterglow searches), and would have detected the host if it were very 
bright in the submm (cf., Smail \etal\ 2002), the few sub-mm detections of GRB 
host galaxies to-date (e.g., Barnard \etal\ 2003; Berger \etal\ 2003; Tanvir \etal\ 
2004) have been at the 3~mJy level (at 850~$\mu$m), and so would not 
have been detected at these limits. 
 
\subsection{Infrared observations}\label{sec:ir} 
 
Observations at the KPNO  
2.1~m telescope were undertaken using the SQIID (Simultaneous Quad 
Infrared Imaging Device) which provides 4 simultaneous colours (J,H,K \& L).  
The sensitivity in the nIR is significantly greater than 
that in the mid-infrared (L-band) and hence only the J, H and K images 
were used for our analysis. A first epoch of observations was obtained  
5 hours after the burst, with a second taken after 9 hours.  
 
A PSF matched image subtraction 
was performed (using the ISISII code (Alard \& Lupton 2000)). This 
led to the discovery of a transient source, seen to fade in 
the J, H and K bands. Using the UCAC-2 catalog we find a celestial  
position of  RA$=$11:18:32.63, Dec$=$+15:02:59.9, with a positional accuracy of 
0\farcs1. The discovery images are shown in Figure~\ref{fig1}. 
 
Further IR observations were taken with INGRID at the William Herschel Telescope  
(WHT), and the NOTCAM on the Nordic Optical Telescope, both 
sited on La Palma, SOFI on the 3.6~m New Technology Telescope 
at the European Southern Observatory (ESO) La Silla site (J,H and 
K-band, 26 hours after burst), and finally at the Very Large Telescope 
(VLT, UT1) at ESO, Paranal, using the ISAAC array.  
All of these IR observations were  
reduced through the {\ORAC} pipeline (Cavanagh \etal\ 2003). 
A summary is given in Table~1.

\begin{deluxetable*}{lllrrr} 
\tablecolumns{5} 
\tablewidth{0pt} 
\tablecaption{Ground based optical and nIR photometry of the afterglow of GRB~030115} 
\tablehead{\colhead{\delt (days)} & \colhead{Inst./Filter} & 
\colhead{seeing (\arcsec)} & \colhead{mag}& \colhead{$F_{\nu}$ Jy} & \colhead{$F_{\nu}$ (Host Subtracted)} }  
\startdata 
0.11 & Asiago/1.82m/R & 2.5 & 21.8 $\pm$ 0.3 & 5.5 $\times 10^{-6}$ &5.3 $\times 10^{-6}$\\ 
0.21 & KPNO2.1m/J &1.2&19.19 $\pm$ 0.11 & 3.31 $\times 10^{-5}$ & 3.20 $\times 10^{-5}$\\ 
0.21 & KPNO2.1m/H &1.2&17.82 $\pm$ 0.09 & 7.52 $\times 10^{-5}$ & 7.19 $\times 10^{-5}$\\ 
0.21 & KPNO2.1m/K &1.2&16.84 $\pm$ 0.05 & 1.19 $\times 10^{-4}$ & 1.14 $\times 10^{-4}$\\ 
0.38& KPNO2.1m/J &1.3&20.47 $\pm$ 0.27 & 1.02 $\times 10^{-5}$  & 9.10 $\times 10^{-6}$\\ 
0.38 & KPNO2.1m/H&1.3 &18.37 $\pm$ 0.16 & 4.53 $\times 10^{-5}$ & 4.19 $\times 10^{-5}$\\ 
0.38 & KPNO2.1m/K&1.3 &17.26 $\pm$ 0.08 & 8.12 $\times 10^{-5}$ & 7.65 $\times 10^{-5}$\\ 
1.11 & WHT/H 	  &1.3& 20.29 $\pm$ 0.28 & 7.44 $\times 10^{-6}$& 4.10 $\times 10^{-6}$\\ 
1.22  & NTT/J    &0.7  &21.12 $\pm$ 0.19  & 5.59 $\times 10^{-6}$ &4.49 $\times 10^{-6}$ \\ 
1.23  & NTT/H    &0.7  &20.69 $\pm$ 0.31  & 5.35 $\times 10^{-6}$ &2.01 $\times 10^{-6}$\\ 
1.25  & NTT/K    &0.7  &19.51 $\pm$ 0.15  & 1.02 $\times 10^{-5}$ &5.50 $\times 10^{-6}$\\ 
1.14  & VLT/FORS2/I &0.8&23.8 $\pm$ 0.20   & 8.00 $\times 10^{-7}$&4.02 $\times 10^{-7}$\\ 
1.17  & VLT/FORS2/R &0.8&24.5 $\pm$ 0.20   & 5.49 $\times 10^{-7}$ &$1.89\times 10^{-7}$\\ 
1.21  & VLT/FORS2/V &0.8&25.1 $\pm$ 0.30   & 3.88 $\times 10^{-7}$&1.68 $\times 10^{-7}$\\ 
2.10  & NOT/NOTCAM/K & 1.0& 19.8 $\pm$ 0.3 & 7.24 $\times 10^{-6}$ &2.54 $\times 10^{-6}$\\ 
2.12  & VLT/ISAAC/K &0.7& 19.8 $\pm$ 0.15& 7.24 $\times 10^{-6}$ & 2.54 $\times 10^{-6}$ \\ 
2.20  & VLT/FORS1/I &0.8& 24.0 $\pm$ 0.3 & 7.24 $\times 10^{-7}$ &-\\ 
2.24  & VLT/FORS1/R &0.8& 25.2 $\pm$ 0.4& 2.40 $\times 10^{-7}$ &-\\ 
\hline 
{\bf GCN data} & Kato \etal\ &2003a,b\\ 
\hline 
0.86 	& IRSF/J &	& 20.4 $\pm 0.2$ & $1.16 \times 10^{-5}$ &1.05 $\times 10^{-5}$\\ 
0.86	& IRSF/H &	& 19.9 $\pm 0.3$ & $3.90 \times 10^{-5}$ &3.56 $\times 10^{-5}$\\ 
0.86	& IRSF/K &	& 18.5 $\pm 0.2$ & $2.44 \times 10^{-5}$ &1.97 $\times 10^{-5}$\\ 
1.96	& IRSF/J &	& 21.5 $\pm 0.5$ & $4.2 \times 10^{-6}$  &3.10 $\times 10^{-6}$\\ 
1.96	& IRSF/H &  	& 20.4 $\pm 0.4$ & $6.7 \times 10^{-6}$  &3.30 $\times 10^{-6}$\\ 
1.96	& IRSF/K &       & 19.1 $\pm 0.1$ & $1.4 \times 10^{-5}$ &9.30 $\times 10^{-6}$\\ 
\enddata 
\tablecomments{Table of observations of GRB~030115. Shown is the time since burst 
of each observation, the measured magnitude, flux and host subtracted 
flux. For the final two R and I band points, the 
flux levels are nominally above those measured for the host galaxy, but consistent within 
the measurement errors. All of  
the data points are corrected for foreground reddening E(B-V) = 0.024, assuming 
a MW extinction law.} 
\label{tab:ground} 
\end{deluxetable*}

\subsection{Optical Observations of the GRB~030115 afterglow} 
 
Optical observations of the GRB~030115 error box were performed by 
several groups. However these observations failed to reveal any 
possible candidates in comparison with the Digitized Sky Survey 
(Castro-Tirado \etal\ 2003; Masetti \etal\ 2003; Flaccomio \etal\ 2003; 
Lamb \etal\ 2003). We obtained data with the 1.82~m  
Copernico Telescope (first described by Masetti \etal\ 2003).  
at \delt =2 hours in poor seeing ($\sim 
2.5$\arcsec). The images were bias subtracted and flat fielded in {\IRAF}  
using standard methods. The data contain a weak ($4\sigma$) detection 
of the transient at this epoch from which we determine 
R$=21.8\pm0.3$.  
 
Further optical data were obtained at the VLT using the FORS2  
instrument in V, R and I on 2003 January 16 (\delt =26 hours) 
and in R and I on 2003 January 17, using FORS1. These data were also reduced 
with {\IRAF}, and again summarised in table~1.

 \subsection{{\it HST} observations of the GRB~030115 host galaxy} 
 
GRB~030115 was first imaged with the {\it Hubble Space Telescope (HST)}  
on 2003 February 10. At this 
epoch we obtained images in F606W (broad V-R) and F814W(I) filters using the  
Advanced Camera for Surveys (ACS) Wide Field Camera (WFC) as well 
as in F160W(H) using NICMOS and the NIC3 camera. A second epoch of 
observations were acquired on 2003 April 24, 
in F814W(I) and F110W (J) filters.  
A final set of observations were obtained on 2003 June 16 and 20 in  
F435W (B) and F222M (K) filters, providing us with a complete set 
of B(V-R)IJHK photometry on the host galaxy of GRB~030115.  
The results of these observations are summarised in table~2
(a subset of these observations have also been presented by 
Dullighan et al. 2004).

\begin{deluxetable*}{lllrr} 
\tablecolumns{5} 
\tablewidth{0pt} 
\tablecaption{Optical and nIR observations of the GRB~030115 host galaxy.} 
\tablehead{\colhead{\delt (days)} & \colhead{Inst./Filter} & \colhead{Exptime (s)} &
\colhead{mag}& \colhead{$F_{\nu}$ (Jy)}}  
\startdata 
26.466  & {\it HST}/ACS/F606W 		& 2000	&	25.58 $\pm$ 0.05	& $2.00 \times 10^{-7}$\\ 
26.406 	& {\it HST}/ACS/F814W 		& 1600 	&	24.83 $\pm$ 0.04	& $4.06 \times 10^{-7}$\\ 
26.268  & {\it HST}/NICMOS/F160W	& 5120 	&	22.68 $\pm$ 0.08	& $2.55 \times 10^{-6}$\\ 
126.193 & {\it HST}/NICMOS/F110W	& 2558 	&	24.08 $\pm$ 0.11  	& $1.10 \times 10^{-6}$\\ 
127.672 & {\it HST}/ACS/F814W 		& 1920  &	24.85 $\pm$ 0.04 	& $3.98 \times 10^{-7}$\\ 
151.975 & {\it HST}/ACS/F435W 		& 8800  &	25.99 $\pm$ 0.12  	& $1.37 \times 10^{-7}$\\ 
156.071 & {\it HST}/NICMOS/F222M	& 7678  &	22.21 $\pm$ 0.15  	& $4.70 \times 10^{-6}$\\ 
\hline 
\enddata 
\tablecomments{Table of observations of the host galaxy 
GRB~030115 obtained with {\it HST}. The magnitudes have been 
calculated via SExtractor and are given as AB-magnitudes. 
They have been corrected for foreground extinction assuming 
E(B-V) = 0.024, and a MW extinction law. The magnitudes reported 
are those measured in an aperture determined by a SExtractor  
detection on the F160W image, then applied to other images which 
have been matched to this.}
\label{tab:ground} 
\end{deluxetable*}

The ACS data were ``On-The-Fly'' calibrated to produce flattened 
images. They were then drizzled (Fruchter \& Hook 2002) onto an 
output grid with pixels 0.66 times the size of the native ACS pixel 
(this corresponds to $\sim  0\farcs 033$). The linear drop size ({\tt 
pixfrac}) was set to one.  
 
In order to estimate the contribution of the afterglow to the late 
{\it HST} observations  
we aligned and subtracted the two F814W epochs taken at \delt =26 and 127 days 
respectively.  
Our VLT/ISAAC image, taken 2 days after the burst, was also 
aligned with the F814W image, using 22 point sources common 
to each frame, allowing us to locate the afterglow position to 
$\sim 0\farcs05$ accuracy. 
At this position we determine a $3\sigma$ upper limit  
for the afterglow flux of F814W(AB) $= 28.2$ (at \delt =26 days).  
Based on the measured 
R-band magnitude of R$=21.8$ at \delt =2 hours and the  
optical slope of $\alpha = -1.4$ we would predict that the 
afterglow should have had $I \sim 31$ at the time of this observation 
and thus the non-detection is in accordance with what we  
would expect by extrapolation of the early decay. We conclude that all of 
our {\it HST} observations contain negligible afterglow contamination of the host. 
 
An additional possible cause of contamination could occur due to  
a possible supernova. Supernovae spectra peak in the 
optical rest frame, which lies in the IR bands at our favoured 
redshift (see Section 3.1). We do not have multi-epoch IR observations 
to test directly via image subtraction, however the expected flux 
of even an extreme event like SN~1998bw at $z\sim2$ is $<0.05$ $\mu$Jy  
(F160W(AB) $>$ 27). We therefore do not believe that there could 
be a visible supernova component in our first IR images.

\subsection{Photometry and Light Curve} 
 
We calibrated each of our ground based nIR fields by use of  
secondary standards within the field of view of all telescopes. These  
stars are shown in Table~3. We performed aperture photometry on  
the standard stars and the afterglow candidate with the aperture 
being set to twice the Full Width Half Maximum (FWHM) of the  
image. The RMS scatter observed in the magnitudes of  
the standard stars in each of the observations (from that expected and  
tabulated in Table~3) is less than 0.06 magnitudes. 
 
The lightcurve of GRB~030115 in J, H and K is shown in Figure 2.  
Fitting the three bands simultaneously yields a power-law
decay slope of $\alpha = 1.26 \pm 0.06$. However a single
power-law is a relatively poor fit to the data, largely due
to the second epoch J-band observation, which falls signficantly below
the expectation of the power-law at this time.

 Extrapolating our R-band observations (R=21.7 at 0.11 days) to the time of the 
first IR epoch (assuming $t^{-1.26}$ as the decay slope) yields R=22.6, which 
results in an R-K colour of 5.7, making GRB 030115 one of the most reddened 
afterglows observed to date. 
 However it should be noted that this extrapolation
assumes that the afterglow behaviour is predictable over this time, the 
possible unusual behaviour of the J-band lightcurve and the frequent presence
of bumps in the afterglow light curves of many GRBs mean that this value should
be used with caution.

\begin{deluxetable*}{llrrr} 
\tablecolumns{5} 
\tablewidth{0pt} 
\tablecaption{Secondary stars used for photometric calibration in the GRB 030115 field} 
\tablehead{\colhead{RA($^{\circ}$)} & 
\colhead{Dec$^{\circ}$}& \colhead{J} & \colhead{H} &  \colhead{K}}  
\startdata 
169.65241& 15.06268& $ 16.13 \pm 0.02$&  $15.63 \pm 0.03$&  $15.44 \pm 0.10$\\ 
169.64557& 15.05777& $ 15.84 \pm 0.03$&  $15.45 \pm 0.01$&  $15.36 \pm 0.06$\\ 
169.65432& 15.04912& $ 14.79 \pm 0.02$&  $14.15 \pm 0.02$&  $13.98 \pm 0.02$\\ 
169.61094& 15.04204& $ 15.93 \pm 0.05$&  $15.44 \pm 0.04$&  $15.08 \pm 0.06$\\ 
169.62917& 15.03929& $ 15.61 \pm 0.05$&  $15.17 \pm 0.02$&  $15.13 \pm 0.06$\\ 
\enddata 
\label{tab:ground} 
\end{deluxetable*}

\section{The Host Galaxy -- Photometric Redshift} 
 
The host galaxy of GRB~030115 (Fig.~4) shows an irregular 
morphology. The host may be involved in a merger with a companion galaxy offset 
approximately 1\arcsec\ northeast of the host. The F435W image (the 
shortest wavelength used in this study) also shows evidence for  
an edge-on 
disk through the host galaxy. Our precise ISAAC astrometry allows us to 
position the afterglow at the southern end of this putative disk, with a positional 
accuracy of $0\farcs 05$ ($1\sigma$). The host was observed with a complete filter 
set (B(V-R)IJHK) using both ACS and NICMOS. The results of photometry are shown 
in Table~1. The host has a very red R-K color (5.35), far redder than any other GRB 
host observed to-date. This color is dominated by a break between the J and H bands of 
1.4 magnitudes (AB). It is possible that this represents the Balmer (4000 \AA~) break,
which could indicate the presence of an underlying older stellar population, however it
is not possible to distinguish unambiguously between an strong break as might be 
expected from an older population and a very steep slope which could be indicative of a 
younger, but obscured system.

Knowledge of the redshift of GRB~030115 is essential to our understanding of the 
afterglow, and hence to the issue of ``dark'' bursts. Despite attempts from the 
ground to obtain spectroscopy of the host galaxy, the results were inconclusive. We 
have therefore sought to determine a photometric redshift from our B(V-R)IJHK 
photometry.  In addition we have also determined photometric redshifts for 
other galaxies which fall within the field of view of our ACS and NICMOS  
images (the complete region has an area of 0.51 arcmin$^{2}$) for 
a complete set of filters. 
 
We have applied two independent codes. The first is the template fitting method 
of {\hyperz} \footnote{see {http://webast.ast.obs-mip.fr/hyperz/}}; the 
second is the Bayesian method of Ben\'{\i}tez (2000) \footnote{see 
{http://adcam.pha.jhu.edu/$\sim$txitxo/bpzdoc.html}}. They are outlined below but 
for more detail the reader is referred to Ben\'{\i}tez (2000) (\BPZ) and Bolzonella, 
Miralles \& Pello (2000) ({\hyperz}). 
 
{\hyperz} is a template fitting photometric redshift code. It provides a 
maximum likelihood redshift by performing $\chi^2$ fitting of the available 
photometry with a variety of spectral templates: Starburst (Stb), Elliptical 
(E), Lenticular (S0), Irregular (Im) and a set of spiral templates of type Sa 
through Sd. In addition to the photometric redshift and extinction measures it 
is also possible to obtain an estimate of the age of the stellar population of 
the host galaxy. The impact of the assumed IMF when {\hyperz} has been applied to 
host galaxies has been negligible (Gorosabel \etal\ 2003a,b; Christensen  
\etal\ 2004). 
 
The Bayesian code (\BPZ) works by making use of other available information 
information, 
rather than simply fitting templates. These priors take into account our 
knowledge of the expected physical parameters of galaxies, such as their 
luminosity function, redshift distributions, and galaxy type fractions. Here we 
use a luminosity prior based on the observed luminosity function seen in the 
Hubble Deep Field. Therefore, in addition to the template fitting performed by 
{\hyperz}, {\BPZ} also determines the probability that a galaxy of that 
magnitude should be present at the redshift suggested by the linear fit. This 
can be effective at clipping ``unreasonable'' redshifts when template fitting 
alone finds two or more plausible fits.  
 
Magnitudes for the photometric redshifts were determined by aligning 
and rebinning all of our ACS and NICMOS data to the same pixel 
scale, while preserving photometry (this was done via the IRAF 
tasks {\it geomap} and {\it geotran}). We then defined a detection 
mask on the F160W image and extracted magnitudes from the  
same physical region in other filters. We thus obtain a complete 
set of photometry to H=25(AB). The ACS images are significantly  
deeper than those obtained with NICMOS and therefore contain many  
galaxies which are not contained within our catalog. However in 
the absence of IR detections it is not possible to determine  
accurate photometric redshifts for these galaxies.

Based on these detection criteria we determine photometric 
redshifts of $z=2.50 \pm 0.20$ 
for \hyperz\ and $z=2.70 \pm 0.25$ for \BPZ.  
Individually the best fit redshift of the neighbour is found to be 
$z=2.65 \pm 0.30$ based on \hyperz\ and $z=3.0 \pm 0.30$ based on \BPZ. The colors 
of the neighbour galaxy are very blue (F814W-F160W (AB)= 1) and 
are typical of star forming galaxies containing little dust.   
The best fitting spectral type of the host galaxy is that of a S0 in 
\hyperz and a Sb/c in \BPZ. \hyperz\ finds that the internal  
extinction within the host is moderately large $A_V = 1.0$ and 
that its dominant stellar population is old; 2.6~Gyr. This age 
determination is largely due to the strength of the Balmer break  
which lies between the J and H bands at this redshift. If this 
age determination were correct it would imply a redshift of formation 
of $z\sim 6.5$. However, this age should be treated with caution since 
it has been noted that for Extreme Red Objects (EROs) the photometric shape alone cannot 
adequately distinguish between dusty, star forming galaxies and those 
which harbor an older population of stars (Moustakas \etal\ 2004).  
 
\section{Comparison of the host properties with other GRB host 
galaxies} 
 
Given the very red color of both host and afterglow, it is obviously interesting 
to compare GRB~030115 with other host galaxies. Chary \etal\ (2002) and Le 
Floc'h \etal\ (2003) have imaged a sample of GRB host galaxies in the IR 
(K-band). Their measurements allow the determination of the global (R-K) color 
of the host galaxies and provide insight into the extinction and star formation 
rate. A important question relating to the nature of GRB hosts is whether these 
are typical of high-z galaxies which are selected by other techniques 
(e.g., sub-mm surveys). These surveys indicate that the majority of high-z star 
formation ($z \sim 2$) is obscured with a significant fraction likely 
taking place in galaxies analogous to the lower redshift ULIRGs. The optical counterparts to these sub-mm bright galaxies are 
often very red, massive systems. By contrast, typical GRB host galaxies are 
mostly 
blue and sub-luminous, though the latter may be due to the steep 
luminosity function of star-forming galaxies (Fruchter \etal\ 1999; Hogg \& Fruchter 1999;
Fynbo \etal\ 2002; Le Floc'h \etal\ 2003). The R-K 
colors of GRB host galaxies are therefore apparently distinct from those of the 
typically red sub-mm emitting galaxies, and the few GRB host 
galaxies which do show sub-mm emission do not show any significant optical 
properties to distinguish them from other GRB hosts (Berger \etal\ 2003). In fact a sub-mm emitter, 
like the host of the dark GRB~000210, can show low extinction in the optical-nIR 
having colours 
typical of the bluest host galaxies (Gorosabel \etal\ 2003a). 
The host of GRB~030115 is, however, very red and 
lies at a redshift comparable to that of many sub-mm sources ($z \sim 2.4$; 
Chapman \etal\ 2003). Although early sub-mm afterglow observations have  
only provided an upper limit to the 850~$\mu$m flux (Hoge \etal\ 2003), the depth of these observations 
would not have been sufficient to detect any other GRB host to-date 
(Tanvir \etal\ 2004). 
 
Figure 6 shows the R-K color of the host galaxies of GRBs versus their redshift. 
K-band data is taken from Le Floc'h \etal\ (2003) and Chary \etal\ (2002) (and 
references therein). The host galaxy of GRB~030115 lies 
significantly above the curve that represents the majority of the GRB host 
galaxies. Indeed its spectral type is best fit as an Sc; however it remains 
significantly underluminous in the optical (rest-frame UV) (R* / 8).  In contrast 
it is apparently overluminous in the K-band (K* $\times$ 4). It is 
possible that this is due to extinction within the disk of the host 
galaxy which we may view edge on. The majority of the optical light 
produced at the centre of the galaxy could then not escape through the dusty  disk of 
the galaxy, resulting in its red color.  

The absence of a single red sub-mm emitting galaxy from the list of GRB hosts is
particularly puzzling given the widespread belief that most of the sub-mm emission 
from these objects is due to dust heating by massive stars (Le Floc'h \etal\ 2003; Fynbo \etal\ 
2003). The presence of GRB~030115 within a very red host offers some evidence 
that a fraction of GRBs are extincted and come from larger galaxies with red 
global colors; however the proportion of bursts which may lie within this class 
is very poorly constrained so far (see also Klose \etal\ 2000, 2003).  Unfortunately, 
the faintness of the afterglow GRB~030115 in the optical, precluded high-resolution 
spectroscopic study of the ISM of the host. 
 
\subsection{Host Galaxy Star Formation Rate and Extinction} 
 
At $z \sim 2$ our broadband ACS images cover restframe wavelengths of 
approximately 1550, 2160 and 2900~\AA\ for F435W, F606W and F814W, respectively. 
From this we can obtain a measure of the host galaxy extinction and star 
formation rate. 
 
Star formation measures based purely on the rest-frame UV luminosity are prone 
to significant errors due to the effects of dust scattering and absorption of UV 
light. Therefore unextincted measurements provide only a lower limit on the star 
formation rate. Using the relation of Kennicutt (1998), 
\begin{equation} 
\mbox{SFR} (M_{\odot} \mbox{yr}^{-1}) =  
1.4 \times 10^{-28} L_{\nu}  
(\mbox{erg s$^{-1}$ Hz$^{-1}$ (1500-2800~\AA)}), 
\end{equation} 
we obtain an flux (uncorrected for dust extinction) of  
$F_{\nu}$ = $2.13 \times 10^{-30}$~erg~s$^{-1}$~cm$^{-2}$ which, using our  
choice of cosmology, implies a  
value of $L_{\nu} = 4 \pi d_l^2 F_{\nu} /(1+z) = 3.1 \times 10^{28}$~ergs~s$^{-1}$~Hz$^{-1}$.  
This gives a SFR of 4.4~M$_{\odot}$~yr$^{-1}$, in the F606W  
passband.  
 
Given the red colour of the observed afterglow we may expect that there is 
significant dust within the host galaxy and therefore substantial scattering and 
absorption of the restframe UV flux. This UV flux is then re-radiated in the far 
infrared. Meurer, Heckman \& Calzetti (1999; hereafter M99) have calibrated this effect 
for local starburst galaxies and find a strong correlation between restframe 
ultraviolet spectral slope and flux at 60~$\mu$m. Under the assumption that the 
relation found for local starbursts holds at higher redshifts (which may be 
expected, see section 2, M99) we can obtain a better measure of the star 
formation rate, and the extinction within the host galaxy. The relation given in 
M99 for extinction is  
\begin{equation} A_{1600} = 4.43 + 1.99 \beta.\end{equation}  
The relation for spectral slope for $(F606W-F814W)_{AB}$ at 
redshift $z$ is given by \begin{equation} \beta = 3.23(F606W - F814W)_{AB} - 
5.22 + 2.66z - 0.545z^2. \end{equation}  
This results in $\beta = 0.44$, corresponding to  $A_{1600}$ = 5.3. Correcting our results to this extinction 
implies a massive star formation rate of $\sim 500 M_{\odot}$~yr$^{-1}$. 
 
While this result remains highly uncertain, taken together with the extreme red  
color of the host galaxy, makes GRB~030115 an excellent  
candidate for further sub-mm followup.

 \section{A possible cluster at $z \sim 2.5$} 
\label{cluster} 
In addition to obtaining the photometric redshift for GRB~030115 we have also been 
able to extract photometric redshifts for other galaxies within the field. 
The ACS observations contain many more objects than the NICMOS data, due
to its greater depth and field of view. However these observations do not provide
sufficient color information to enable photometric redshifts to be accurately 
determined. Thus only objects detected in the F160W image with
magnitudes brighter than F160W(AB) =25.0 are included in 
the photometric redshift fitting. Stellar objects are rejected based on their
extent within the ACS images (which have a much smaller point spread
function that the NICMOS observations). A total of 56 galaxies meet
this requirement. 
The histogram of photometric redshifts determined by  \hyperz\  is shown in Fig~7.
As can  
be seen there exists a strong peak in the redshift distribution at $z=2.5$, 
the favoured photometric redshift for the host galaxy using these selection 
criteria. The RMS of galaxies at redshift $2 < z < 3$ within this fit is 
$\sim 0.12$, indicating that there may be an association (such as an edge-on sheet 
of galaxies) or possibly a cluster at this redshift. The  
galaxies which form this association are all galaxies with luminosity at, or  
somewhat large than L* (in the H-band). We are therefore only seeing the brightest 
members, while the more numerous, fainter galaxies are not 
detected. 
 
In order to quantify the potential significance of this association it is necessary to  
determine the level of overdensity which exists at this redshift. To achieve  
this we need to examine the distribution of redshifts seen in other deep surveys. 
Few  ground based surveys reach the limits obtained here (although most cover a much wider area). 
The deep surveys which we chose to use here were the GOODs survey (Giavalisco \etal\ 2004), 
the Hubble Deep Field North (HDF-N) and South (HDF-S) and the Hubble Ultradeep Field (UDF).  
For the HDF-N and UDF there is deep NICMOS imaging in F110W and F160W, for the GOODs survey 
and the HDF-S the IR imaging comes from ground based observatories, mostly deep 
VLT/ISAAC imaging. The optical imaging in  
the HDF-N and HDF-S is by WFPC2  
and in the case of GOODs and the UDF is with ACS. 
In order to confirm the accuracy of our photometric redshifts determined by \hyperz\ 
we performed fits on each of the photometric catalogs and compared the accuracy  
with known spectroscopic redshifts.  In order to compare our data with what can
be obtained with similar quality data we restricted the catalogs from 
each of these surveys to F160W(AB)$<$25 or H(AB)$<$25 for ground based
observation. We also used the same filter set as obtained for the host of GRB 030115
(in general this means that we did not include U-band observations).

A potential problem with the determination of photometric redshifts via the template 
fitting {\it only} is that at times the redshift is poorly constrained (due often 
to a very low $\chi^2 / dof$ ($<< 1$)). This commonly leads to low redshift 
systems being placed in higher redshift bins. Bayesian techniques can  
solve this problem by eliminating systems which would be overly luminous 
at a given redshift. Using \BPZ\ on both the HDF and the GRB~030115 field 
does reduce the number of objects lying in the $2 < z < 3$ bin. However 
the Bayesian priors are based on observed luminosity functions of field galaxies 
whereas 
galaxies within clusters may be expected to be more luminous. In practice, 
using \BPZ\ on the GRB~030115 resulted in 5 objects being removed 
to a lower redshift bin and a slight broadening of the measured peak. Nonetheless 
the apparent overdensity remains, favoring a cluster at the redshift of the 
GRB host galaxy. Also comparison of the photometric redshifts with 
the known spectroscopic redshifts in the HDF and the GOODs survey 
implies that with our choice of filters and limiting magnitudes $<10$\%  
of low redshift galaxies are placed at $z>2$ by \hyperz. Thus  we  
conclude that at most $\sim 20$ \%  
of the galaxies placed in the 
$2 < z < 3$ bin are falsely located. Even if these are removed there  
remains a substantial overdensity in the expected number of galaxies 
at this redshift, and we conclude that this is most likely a true association, 
and possibly a cluster of which 
the host galaxy of GRB~030115 is a member. 
 
Without spectroscopic redshifts for a number of members we cannot 
determine the nature of the structure which we see unambiguously; it is possible 
that it is a larger scale association with a broad velocity dispersion which 
is not gravitationally bound. However should the depth of this overdensity be 
comparable to its observed width in our HST images, then the association would be 
a cluster comparable in richness to that of Coma (Somerville 2004).   
 
Figure 8 shows the histogram of colours for all galaxies found to lie in 
the redshift range $2 < z < 3$.  While a few can be seen to have red colors comparable 
to the GRB host galaxy, the majority of the objects have much bluer colors, consistent 
with their being star-forming galaxies at this redshift. The host itself is typical of what  
would be expected of a galaxy in the center of a cluster at its redshift: a red 
and luminous galaxy still forming stars but in the process of evolving into a modern day elliptical. 
 
\subsection{The number of density of EROs about GRB~030115} 
 
EROs are generally defined by a color cut at around R-K = 5, although 
various other more stringent 
 cuts eg., R-K$>$5.35, R-K$>$6, are also used. The original definitions for EROs
 were based on the expected colors of primeval elliptical galaxies (Elston \etal\ 1988). 
 In practice EROs consist both of old passive systems (Daddi \etal\  2000) and dusty
 star forming galaxies.

Our deepest IR image, and thus that most suitable to the discovery  
of red objects is F160W (approximately H-band). Thus we chose 
equivalent colors of F814W-F160W $>$ 1.85 (AB), which 
is an effectively identical spectral slope to R-K=5 in Vega magnitudes. In our region (approximately 0.5 arcmin$^2$) 
we find a total of 5 galaxies meeting this criterion. Three of these lie within 
the $z=2.5$ overdensity with the other two having poor photometry (4 rather than 6 colors 
with non-detections at F110W and F222M) fitting better at z$\sim$ 0.7, but also 
having acceptable fits ($\chi^2 / dof < 2$) at $z=2.5$.  
 
We have also searched a larger region surrounding GRB~030115 by using our ACS image 
in conjunction with the K-band image obtained with ISAAC. The GRB still contributes 
some flux in this image and therefore the photometry of the host galaxy is not 
reliable. We determine that there 
are $\sim 10$ EROs within this 3.2 arcmin$^2$ region to a limit of K$<$23 (AB). 
Comparing this with ERO number counts (e.g. Moustakas \etal 2004; Gilbank \etal\ 2003) 
we find that they are overdense by a factor $\sim$ 2. In practice using the 
``average'' number of EROs is somewhat misleading since EROs are often highly 
clustered. Figure 9 shows the color magnitude diagram for the GRB~030115 region.

\section{The origin of the optical faintness of GRB~030115} 
 
\subsection{Extinction} 
 
GRB~030115 exhibits an exceptionally red color ($R-K\approx6$) but was not at 
extreme redshift. This is very unusual for GRB afterglows which 
typically have spectra well described as $\nu^{-1}$, corresponding to $R-K = 
2.9$. Furthermore the color of the GRB~030115 afterglow is not well defined by a 
single power-law, but has significant curvature in the spectral energy 
distribution occuring at shorter wavelengths. The reddening of the afterglow due 
to extinction within the Milky Way is probably small, $E(B-V) = 0.024$  
(Schlegel, Finkbeiner \& Davies 1998), making extinction in the host galaxy the 
most likely explanation. In order to estimate the extinction along the line of 
sight to GRB~030115 we fit the afterglow with an extinction model of the form 
$F_{\nu} = \nu^{\beta} \times 10^{-0.4 A_{\nu}}$. We have
attempted to obtain the extinction at a 
given frequency $A_{\nu}$ based on the fitting models given by Pei (1993),  
relevant for differing dust to gas ratios and metallicities which are 
reflected in the MW, LMC and SMC. These fits result in unphysical parameters for 
the afterglow spectrum (namely $\beta > 0$, and $A_V = 4$, with MW-like 
extinction),  this is true at all redshifts.
 
The precision of the above measurement could be compromised by 
several effects. The first is that while the afterglow is bright (and hence the 
errors small) we only obtain a truly simultaneous SED in three colors (J,H,K) 
while the R-band data must be extrapolated over a factor 2 in time.
The afterglow does not appear to exhibit a ``pure'' power-law 
behaviour and this this extrapolation introduces a larger uncertainty 
into the spectral energy distribution that simply the photometric uncertainty
associated with the measurement. As such deviations from a single 
power-law decline may naturally explain the poor fit obtained for
the extinction laws described above). Furthermore, at later times
(i.e. $>$24 hours) the errors in the individual photometrc points
are large, and also contain a host contribution which further
reduce their accuracy.

A possible physical explanation is that of dust destruction local to 
the GRB. Under many schemes (e.g., Waxman \& Draine 2000; Fruchter, Krolik \& 
Rhoads 2001) dust can be destroyed by the prompt X-ray/EUV emission over a 
distance of several hundred parsecs about the burst site. This dust destruction 
removes the small grain sizes, leaving the larger ones unaffected and will 
change the extinction law to one dominated by larger grains (e.g., Maiolino {\it 
\etal} 2000). The true extinction along our sight line to the GRB is then the 
composite of the dust local to the GRB environment and that at larger distances 
in the host galaxy (which in this case shows evidence of being dusty). The true 
extinction law may possible be very different from those which 
describe sources in the local Universe and this is the likely reason for the 
poor fit of these laws to our data. A final possibility is that the afterglow 
spectrum has some intrinsic curvature, most likely due to the cooling break 
being in the IR-band at the time of the observation; however the apparent color changes 
seen in the second epoch are not easily explained as the cooling break, 
especially since later data are not consistent with this steepened slope.

\subsection{A dark burst caught?} 
 
It is instructive to compare the colors of the host galaxy of GRB~030115 with 
those of the host galaxies of other GRBs, which typically exhibit blue colors 
(e.g., Fruchter \etal\ 1999). Fig.~10 (left) shows the R-magnitude of a selection of 
GRBs plotted against the R$-$K color of their host galaxies. While this plot is 
limited by small number statistics, it can clearly be seen that 
of the GRBs seen in the R-band at \delt =24~hours, GRB~030115 is the faintest, 
and also lies in the most reddened host galaxy. The best fit line is 
shown, however much power is provided to this fit by GRB~030115 and 
it is therefore not possible to conclusively demonstrate 
a real trend whereby dark (faint) GRBs lie in reddened host galaxies. 
 
Fig.~10 (right) shows the colors of GRB afterglows, plotted against the colors of their 
host galaxies. We have plotted the effective power-law index $\beta$ across the 
broadest range of wavelengths available for an individual burst. This plot shows
little evidence for any correlation between host color and afterglow color, however
the extreme nature of GRB~030115 can be clearly seen since both afterglow and
host galaxy are very red.
 
The magnitude of GRB~030115 at \delt =24~hours is the faintest GRB ever detected in 
the optical at this epoch. The magnitude is deeper than the limits placed on 
many other ``dark'' GRBs (see Fig.~11). This implies that GRB~030115 lies in a 
region of parameter space which would have be consistent 
with a dark GRB. 
 
\section{Conclusions} 
 
We have reported the discovery of the afterglow of GRB~030115. Originating from 
a (photometrically determined) 
redshift of $\sim 2.5$, the afterglow was the faintest seen in the R-band after 
24 hours. The decay rate was faster than the mean of most GRBs with  
an average decay rate over the first 24 hours of $\alpha \sim 
1.5$.  The faintness of the 
afterglow in the optical is therefore likely a function of both this rapid decline 
and its extremely red spectrum. 
 
Comparison of both the afterglow and host galaxy with those observed 
for other GRBs implies that the latter is the most reddened of 
any GRB host galaxy for which an afterglow has been seen. This is 
in contrast to some previous cases where red afterglows 
have been observed to lie in blue galaxies (e.g. GRB~000418, Klose  
\etal\ 2001; Gorosabel \etal\ 2003). However none of these 
afterglows is as extreme as that of GRB~030115, which has a flux density below 
the upper limits set for the large majority of dark GRBs at similar epochs.  
It may therefore be that some of the dark GRBs so far observed evolved  
in similar ways to GRB~030115 and were simply missed due to the lack 
of early IR observations. IR-robotic observations (e.g., REM, Liverpool 
Telescope) may be very succesful at locating such afterglows in the future 
and 
will therefore provide valuable statistics on the number of GRBs occuring 
in obscured galaxies.  In turn these GRB hosts  offer a useful tool in  
characterising the nature of embedded star formation. Indeed moderate 
resolution spectroscopy of these sources at early times may be one 
of the best means of providing a detailed study of the environments 
of such galaxies, which are too optically faint for normal 
absorption line studies. 
 
GRB~030115 also lies in a potentially significant overdensity of galaxies 
at $z \sim 2.5$. This association consists of a high ERO density, consisting
of either dust enshrouded star formation or old, passive systems as well 
as bluer systems with active unobscured star formation. This association 
represents one of the richest associations found at $z>2$. 

However, further observations are necessary to confirm 
the existence of this cluster. Thus far, we have found low-redshift GRBs in field galaxies, however, we expect that as we study GRBs at higher redshifts, we should begin to find them, like GRB~030115, in clusters or in far richer regions of the Universe directly associated with star formation.

\section*{Acknowledgements} 
We thank the referee for an very constructive report which greatly 
enhanced this paper.
Support for Proposal number GO 9405 was provided by NASA through a grant 
from the Space Telescope Science Institute, which is operated by the Association  
of Universities for Research in Astronomy. Incorporated under NASA contract NAS5-26555. 
This work was conducted in part via collaboration within the the 
Research and Training Network ``Gamma-Ray Bursts: An Enigma and a 
Tool'', funded by the European Union under contract number 
HPRN-CT-2002-00294. AJL acknowledges the receipt of a PPARC  
studentship. JMCC 
acknowledges the receipt of a FPI doctoral fellowship from Spain's 
Ministerio de Ciencia y Tecnologia. 
This work was supported by the Danish Natural Science Research Council 
(SNF) and partly based on observations made with the Nordic Optical Telescope, 
operated on the island of La Palma jointly by Denmark, Finland, Iceland, 
Norway, and Sweden, in the Spanish Observatorio del Roque de los 
Muchachos of the Instituto de Astrofisica de Canarias.

\begin{figure*}[ht] 
\centerline{ 
\psfig{file=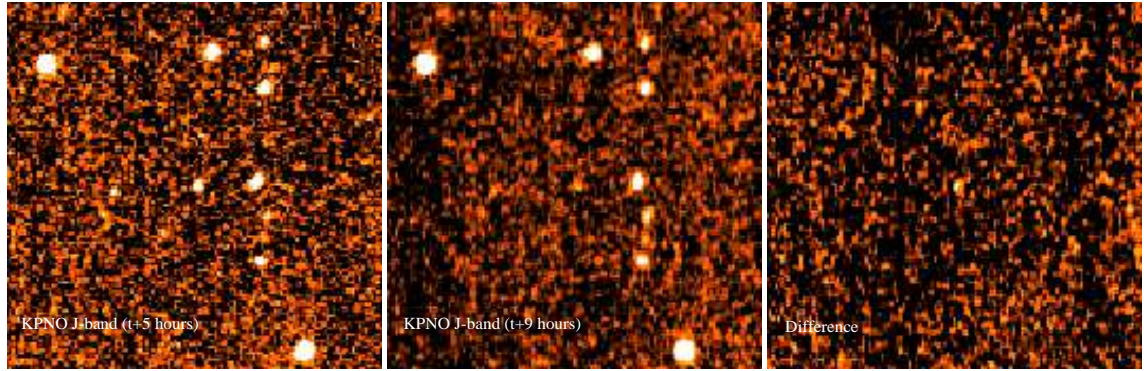,width=6.0in,angle=00}} 
\caption[]{The discovery of GRB~030115: The left panel shows an image 
obtained with KPNO 2.1 + SQIID in the J band 5 hours after burst.  
The central shows an image obtained in the same configuration after 9 
hours, and the right hand panel shows the results of a PSF matched  
image subtraction, the afterglow is at the centre of the image.} 
\label{fig1} 
\end{figure*}

\begin{figure*}[ht] 
\centerline{ 
\psfig{file=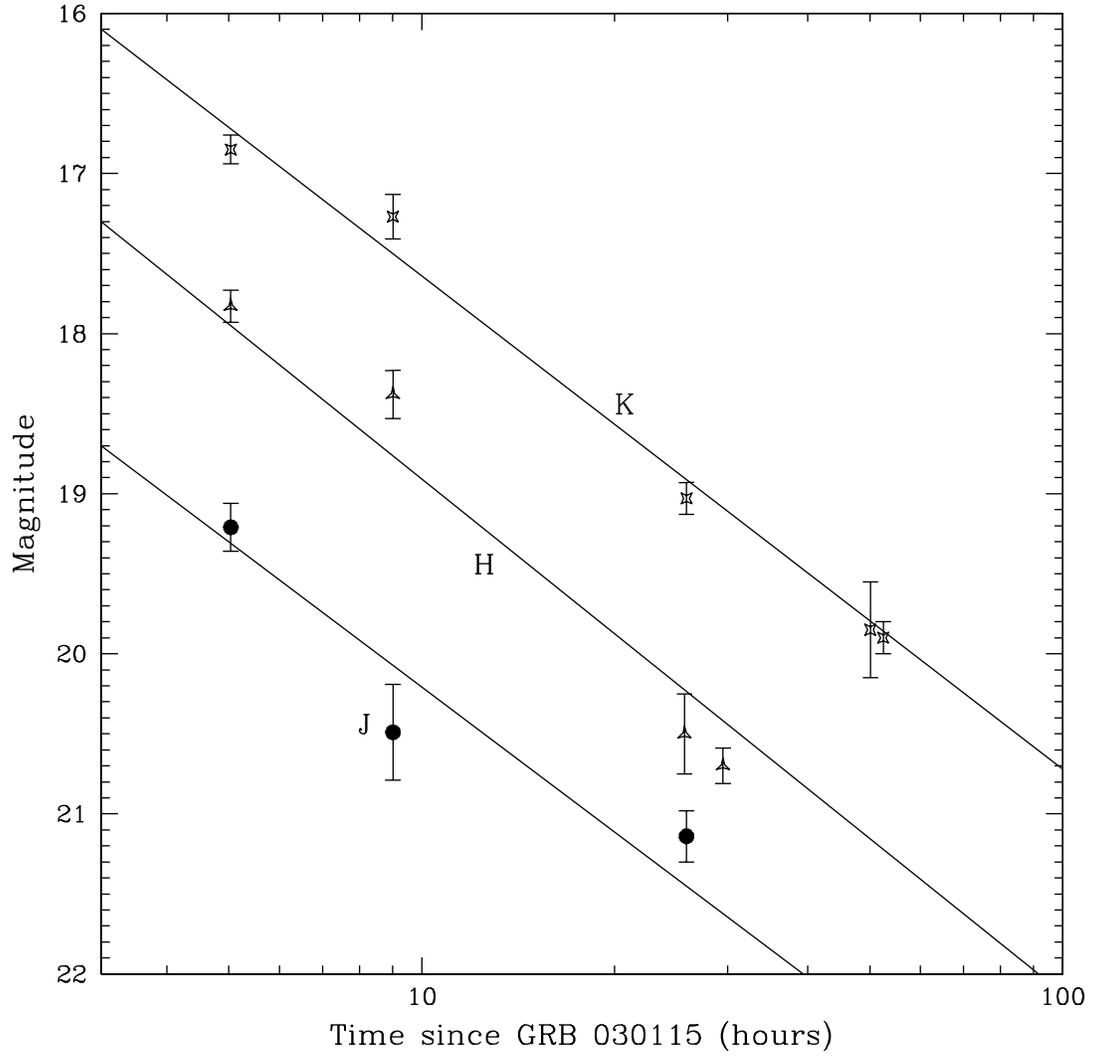,width=6.0in,angle=00}} 
\caption[]{The nIR light curve of GRB~030115 in J, H and K. A broken power-law 
fit is shown and is a good fit to the H and K band light curves. The J band lightcurve 
does not fit this model due to the \delt = 9 hour point.} 
\label{fig2}  
\end{figure*}

\begin{figure*}[ht] 
\centerline{ 
\psfig{file=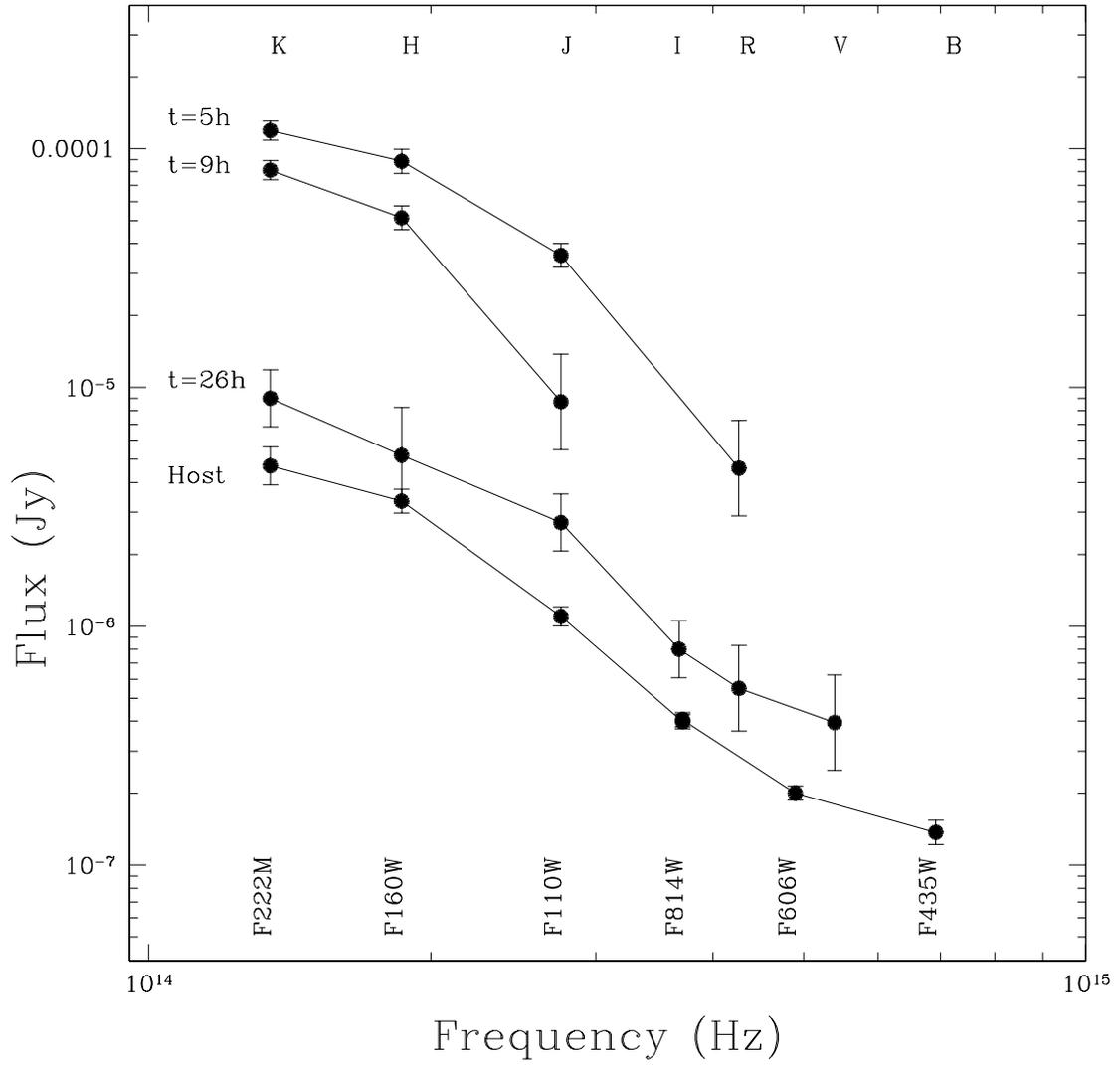,width=6.0in,angle=00}} 
\caption[]{The evolution of the spectral energy distribution (SED) of  
GRB~030115 with time, marginal evidence of color evolution can  
be seen between 5 and 9 hours however the reality of this cannot be 
assessed since it is apparent in only one data point.  The fluxes
shown have been corrected for the effects of galactic extinction.} 
\label{fig3} 
\end{figure*} 
 
\begin{figure*}[ht] 
\centerline{ 
\psfig{file=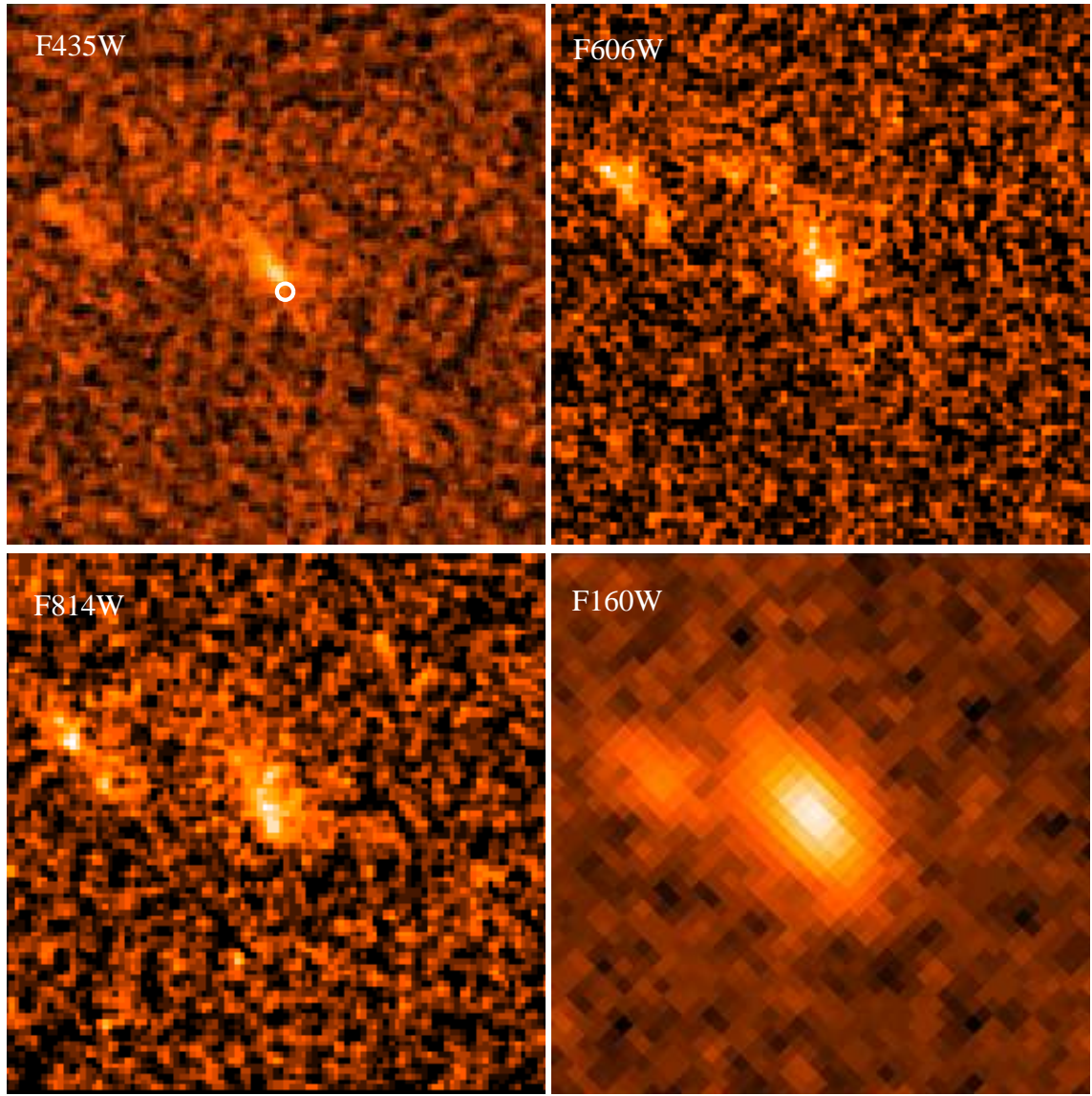,width=6.0in,angle=00}} 
\caption[]{The host galaxy of GRB~030115 as seen by {\it HST}  
using ACS and NICMOS. The position of the afterglow is marked 
as a white circle of the F435W image (top left), with the radius 
of the circle marking the 1$\sigma$ confidence level. The  
possibly interacting galaxy can be seen to the upper left (northeast) 
in each of the images). The field of view of the images is $\sim$ 3 \arcsec} 
\label{fig4} 
\end{figure*}

\begin{figure*}[ht] 
\centerline{ 
\psfig{file=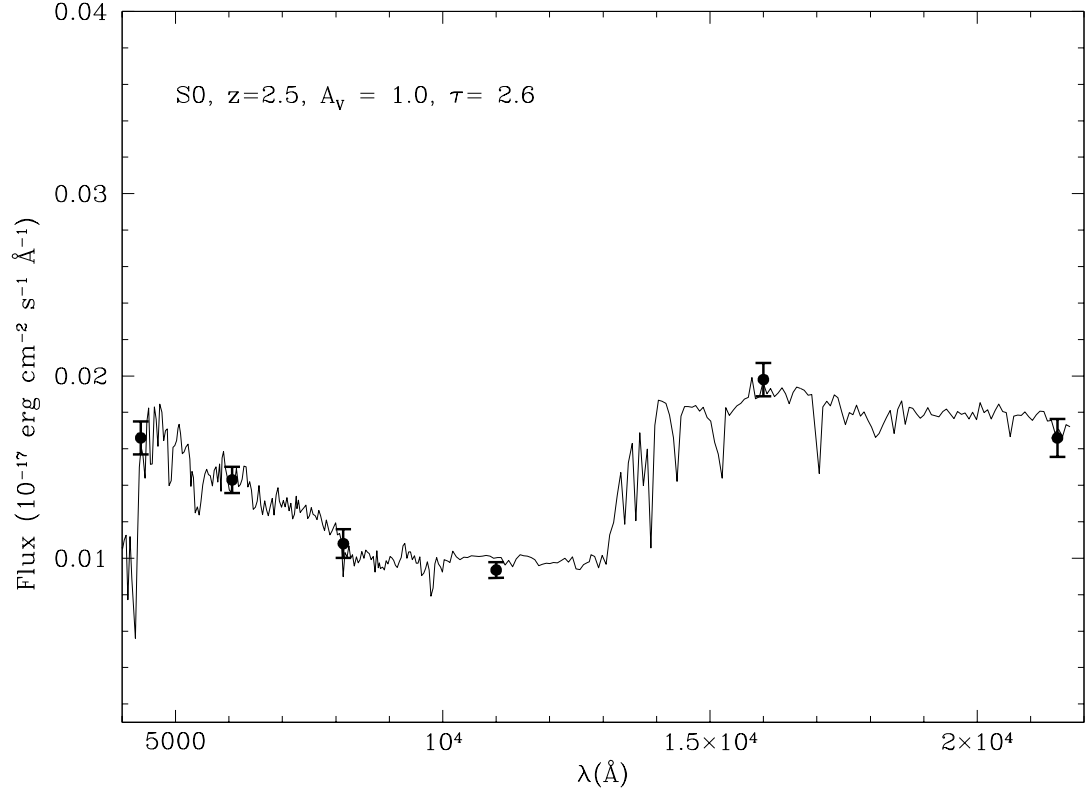,width=6.0in,angle=270}} 
\caption[]{ The best fitting 
spectral energy distribution for the host galaxy of GRB~030115, 
shown as an Sc-galaxy at $z=2.16$, as found by {\hyperz}, 
overplotted are the photometric points from {\it HST}.} 
\label{fig5} 
\end{figure*}

\begin{figure*}[ht] 
\centerline{ 
\psfig{file=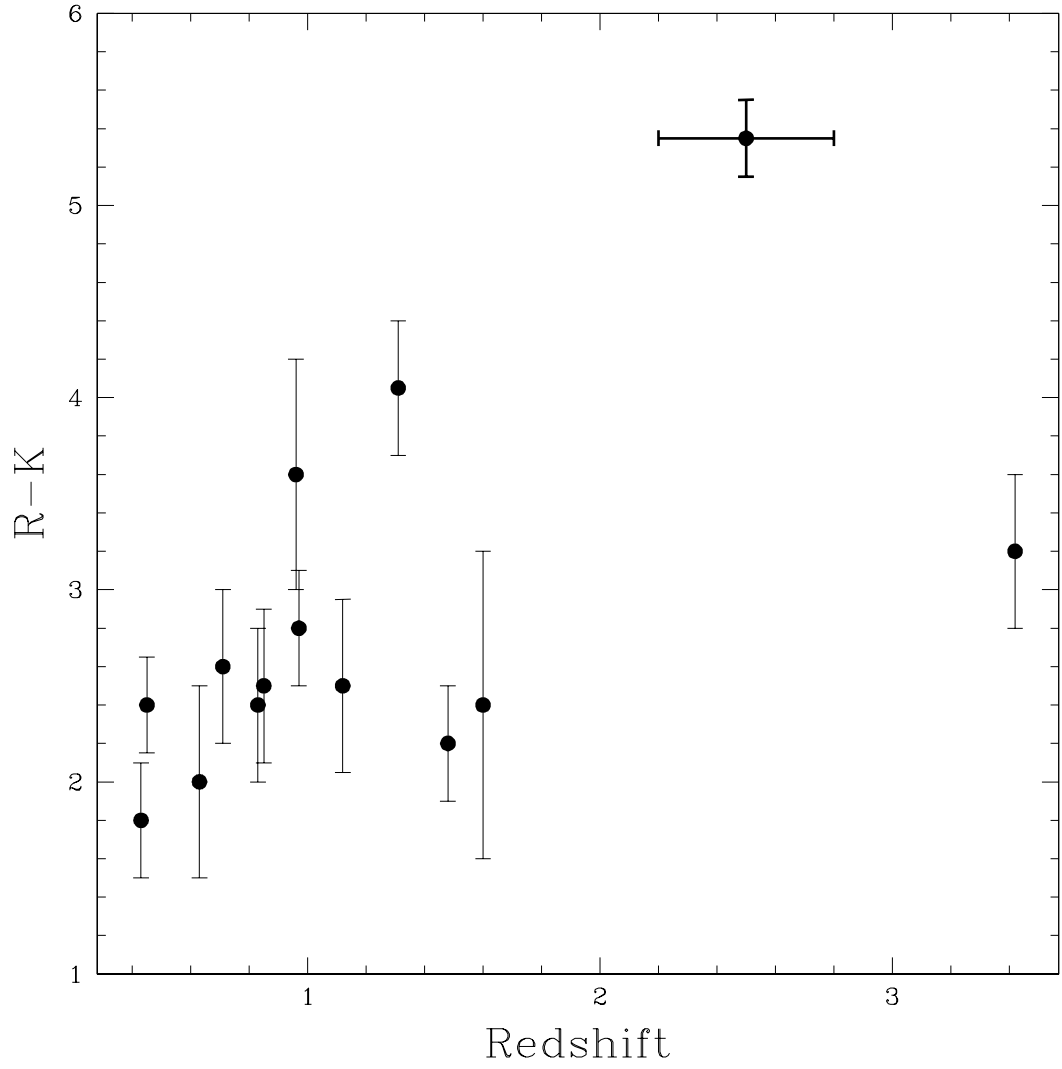,width=6.0in,angle=00}} 
\caption[]{The (R-K) colours of GRB host galaxies as a function of 
redshift (data taken from Le Floc'h \etal\ 2003), the host of 
GRB~030115 is shown in bold (with the errorbar in $z$) it can be seen 
to lie significantly above the typical R-K colors of GRB hosts. } 
\label{fig6} 
\end{figure*} 
 
\begin{figure*}[ht] 
\centerline{ 
\psfig{file=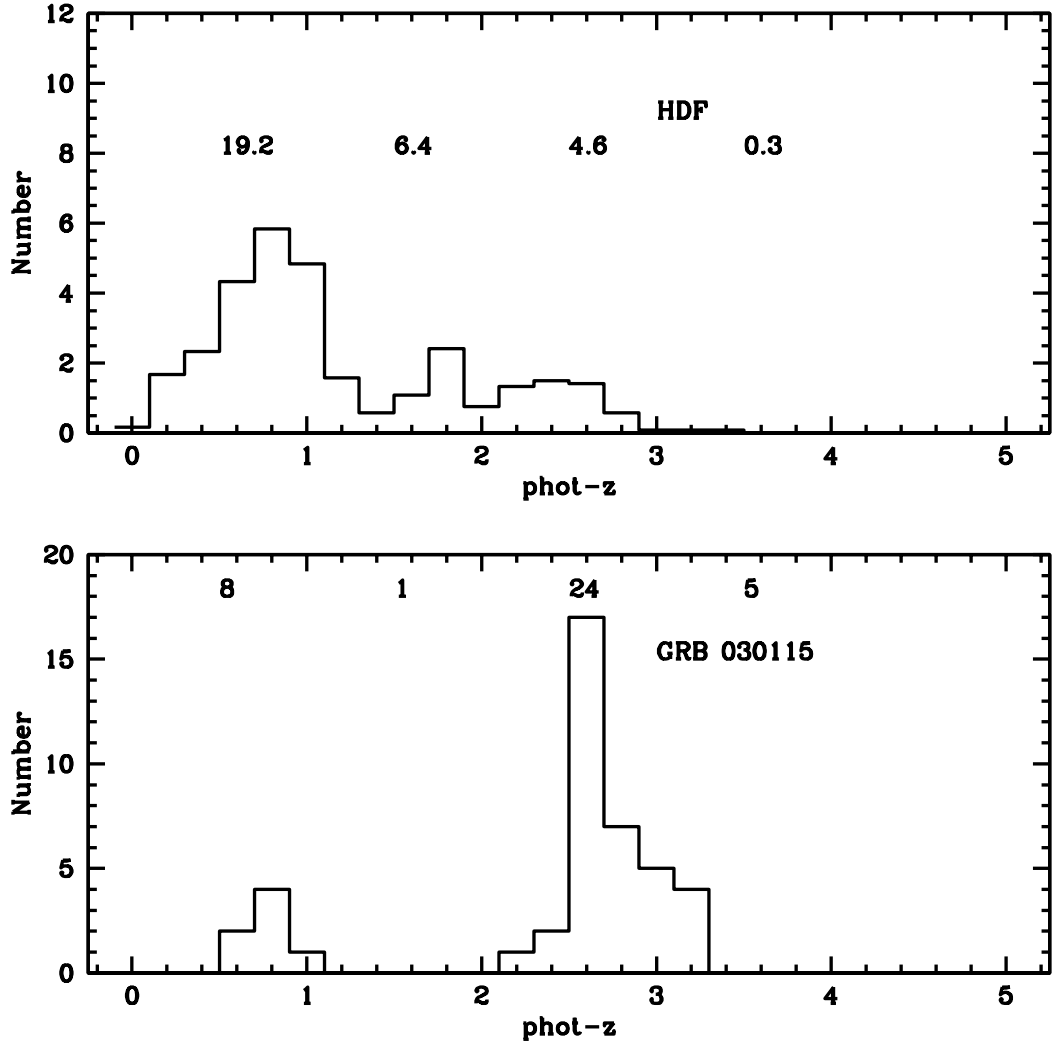,width=6.0in,angle=00}} 
\caption[]{Histogram of photometric redshifts for the HDF (top) and 
the GRB~030115 field (bottom). The HDF photometric redshifts were 
determined from a catalog with F160W(AB) $<25$ and with the 
omission of the U-band data. They have been normalised to  
represent the expected numbers in the GRB~030115 field 
surveyed. The numbers above each histogram represent the 
number of objects found to lie in the redshift bins 
$0 <z<1$,$1<z<2$,$2<z<3$ and $3<z<\infty$ by \hyperz. The possible 
biases within this determination are discussed in Section~\ref{cluster}.} 
\label{phothist} 
\end{figure*}

\begin{figure*}[ht] 
\centerline{ 
\psfig{file=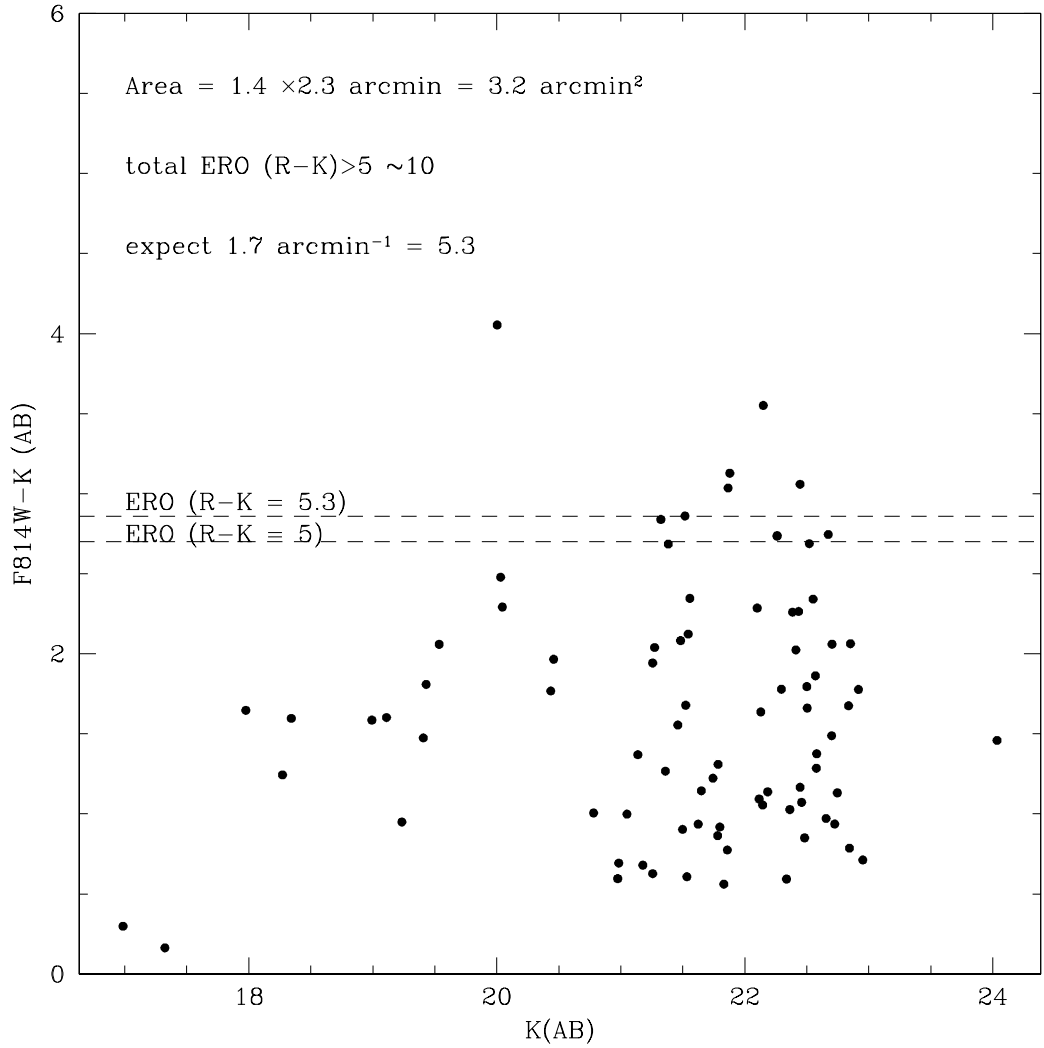,width=6.0in,angle=00}} 
\caption[]{The color magnitude plot for the GRB~030115 field, 
compiled with a combination of ACS and ISAAC data. The color 
cuts which correspond to spectral slopes appropriate 
for R-K =5 and R-K = 5.35 are shown. Approximately  
10 galaxies in this field are EROs to a limiting magnitude 
of K(AB) = 23, there are 
also a number of other galaxies with very red (but not 
extremely red) colors. This is a factor of $\sim$ 2  
overdense with respect to EROs studied within the {\it GOODs} 
field. } 
\label{fig8} 
\end{figure*}

\begin{figure*}[ht] 
\centerline{ 
\psfig{file=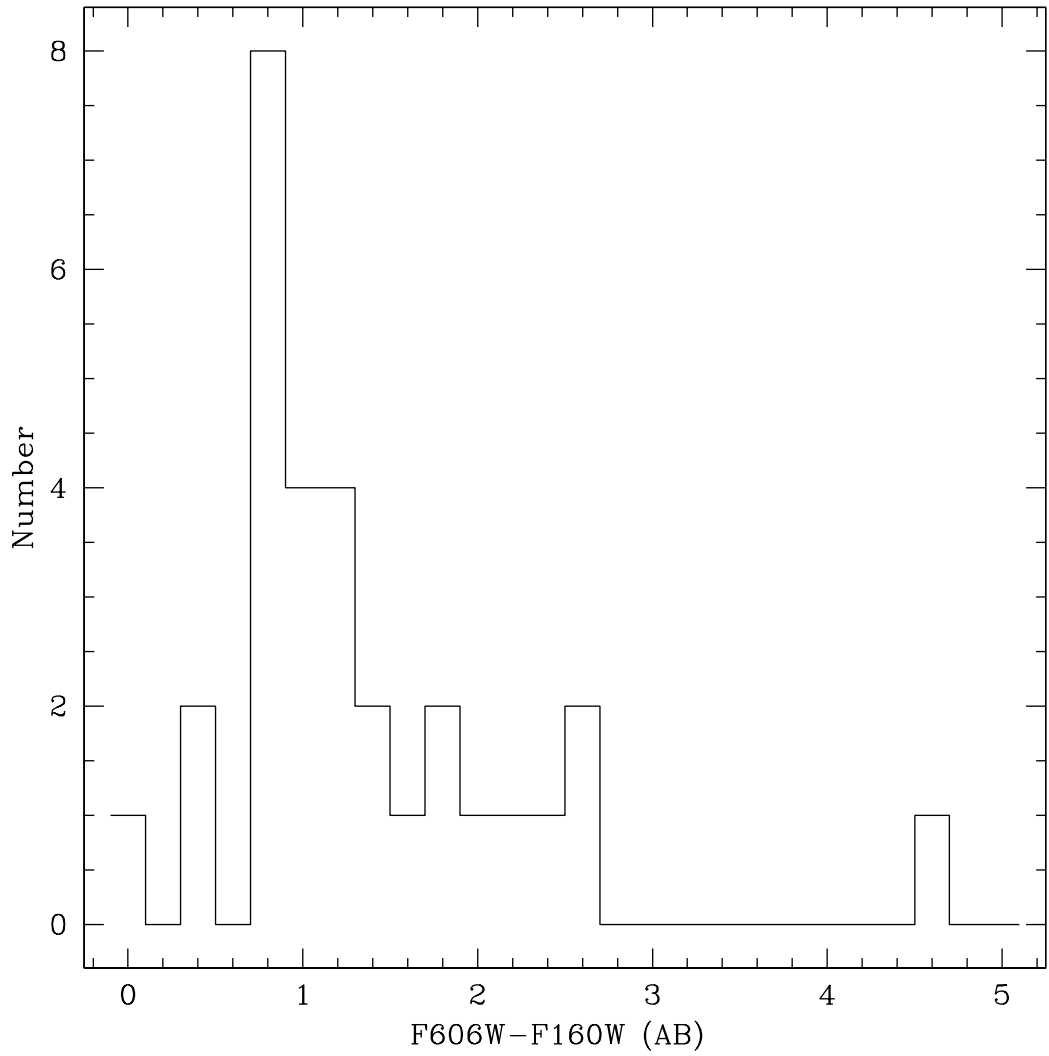,width=6.0in,angle=00}} 
\caption[]{Histogram of the colors of possible cluster members for GRB~030115.  
Although some are very red objects, indicating an older stellar population many 
exhibit blue colors (F606W-F160W(AB) = 1), which are typical of star forming systems. } 
\label{fig9} 
\end{figure*}

\begin{figure*}[ht] 
\centerline{ 
\psfig{file=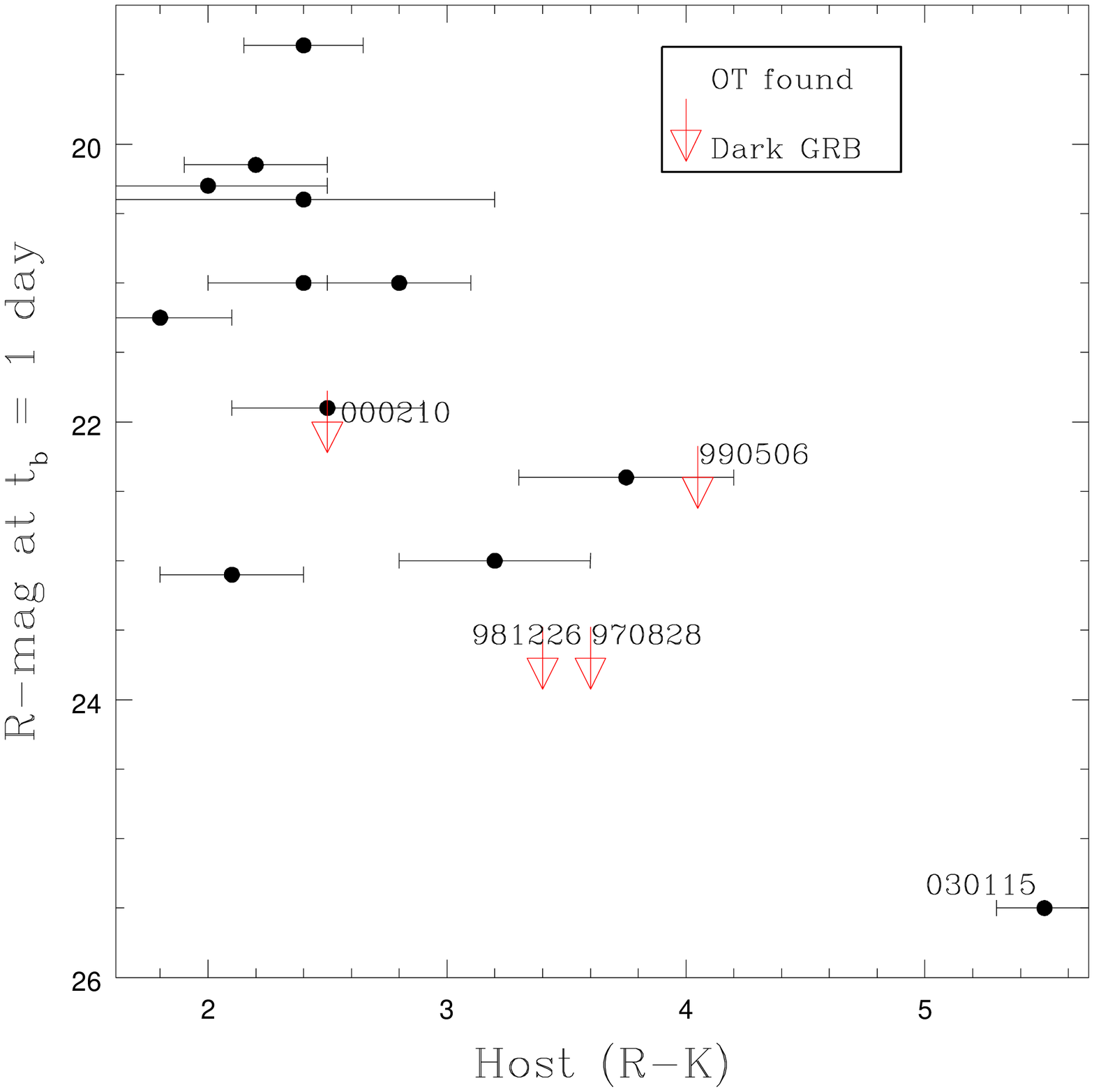,width=3.0in,angle=00}
\psfig{file=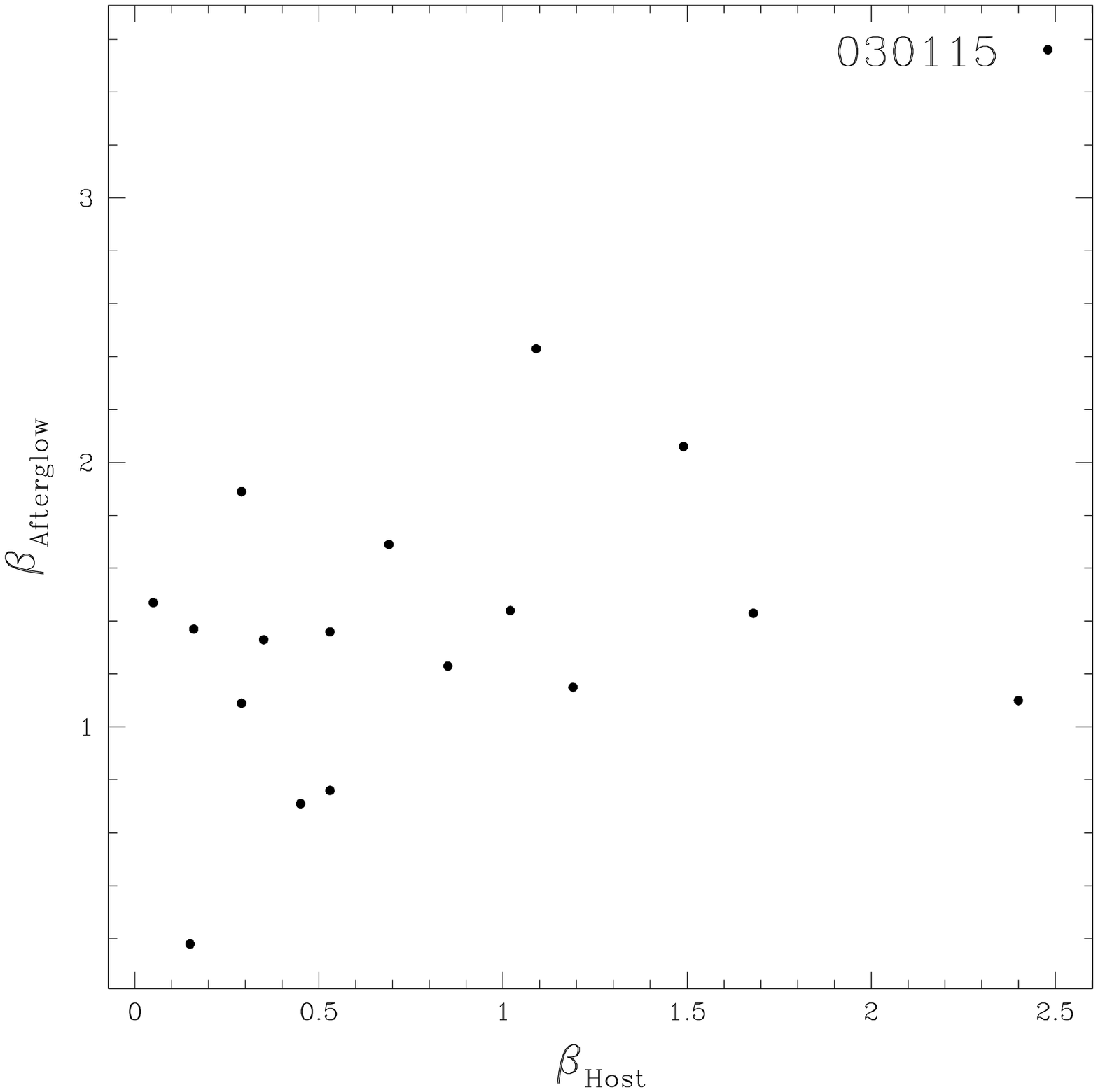,width=3.0in,angle=00}} 
\caption[]{The extreme colors of GRB~030115 and its host galaxy:
{\it left:}
The broadband colours of GRB host galaxies (R-K) versus the  
R-magnitude of their afterglows at \delt =1 day. In cases where 
observations at 1 day are not available the magnitude is calculated 
from an extrapolation of a point near one day, based on the published 
decay slope for each burst. The red arrows indicate the limits placed  
on the dark GRBs. The assumed decay law (for extrapolating the limits 
to \delt =1 day is $\alpha = 1$). The solid line indicates the least 
squares fit to the data. Only those bursts with solid detections have 
been used in the fit. The dark bursts are plotted for comparison.
{\it right:} The colours of GRB afterglows and their host galaxies, plotted  
against one another. The determination of $\beta$ is done on the widest 
wavelength range available for a given afterglow or host galaxy.  
In each case the range of wavelengths used for the determination of  
$\beta$ is the largest available. 
In order to ascertain the colours the magnitudes have been 
normalised to a common epoch using the published values of $\alpha_1$, $\alpha_2$ and 
\delt (the pre and post-break decay slopes, and the time of the break, under 
the assumption that it is the jet break) and assuming an achromatic evolution.  
For low redshift afterglows colour information is taken from the early 
afterglow,where  
supernovae contamination should not affect the observed colour.  
The majority of host colors have been obtained from Le Floc'h et al. 2003; and
Chary et al. 2002. The afterglow colors were obtained from
the following references:
GRB~970228: Galama \etal\ 2000;
GRB~970508: Galama \etal\ 1999;
GRB~971214: Diercks \etal\ 1998;
GRB~980329: Reichart \etal\ 1999, Yost \etal\ 2002;
GRB~980613: Hjorth \etal\ 2002;
GRB~980703: Vreeswijk \etal\ 1999;
GRB~990123: Castro-Tirado \etal\ 1999; Holland \etal\ 2000;
GRB~990712: Sahu \etal\ 2000;
GRB~991208: Galama \etal\ 2000, Castro-Tirado \etal\ 2001;
GRB~000418: Klose \etal\ 2000, Gorosabel \etal\ 2003a (host)
GRB~000926: Harrison \etal\ 2001, Castro \etal\ 2003;
GRB~001011; Gorosbel \etal\ 2002;
GRB~010222 Galama \etal\ 2003;
GRB~011121; Greiner \etal\ 2003;
GRB~020305: Gorosabel \etal\ 2005;
GRB~020405: Masetti \etal\ 2003; 
GRB~020813: Covino \etal\ 2003, Gorosabel \etal\ 2002 (host)}
\label{fig11} 
\end{figure*}

\end{document}